\documentclass[11pt]{article}
\usepackage{gtml2025_workshop}

\usepackage{url} %
\usepackage{graphicx}%
\usepackage{multirow}%
\usepackage{amsmath,amssymb,amsfonts,mathtools}%
\usepackage{amsthm}%
\usepackage{mathrsfs}%
\usepackage[title]{appendix}%
\usepackage{xcolor}%
\usepackage{textcomp}%
\usepackage{manyfoot}%
\usepackage{booktabs}%
\usepackage{algorithm}%
\usepackage{algorithmicx}%
\usepackage{algpseudocode}%
\usepackage{listings}%
\providecommand{\R}{\ensuremath{\mathbb{R}}}
\providecommand{\Z}{\ensuremath{\mathbb{Z}}}

\newcommand{\paren}[1]{\left( #1 \right)}
\newcommand{\brac}[1]{\left[ #1 \right]}
\newcommand{\set}[1]{\left\{ #1 \right\}}

\newtheorem{theorem}{Theorem}
\newtheorem{proposition}[theorem]{Proposition}
\newtheorem{definition}{Definition}

\title{Classification of Histopathology Slides with Persistent Homology Convolutions}

\author{
\textbf{Shrunal Pothagoni}$^{1}$, 
\textbf{Benjamin Schweinhart}$^{1}$ \\[6pt]
$^{1}$George Mason University, Department of Mathematical Sciences, \\ Fairfax, Virginia, USA \\[4pt]
\texttt{spothago@gmu.edu, bschwei@gmu.edu}
}

\date{}

\begin{document}

\maketitle

\begin{abstract}
Convolutional neural networks (CNNs) are a standard tool for computer vision tasks such as image classification. However, typical model architectures may result in the loss of topological information. In specific domains such as histopathology, topology is an important descriptor that can be used to distinguish between disease-indicating tissue by analyzing the shape characteristics of cells. Current literature suggests that reintroducing topological information using persistent homology can improve medical diagnostics; however, previous methods utilize global topological summaries which do not contain information about the locality of topological features. To address this gap, we present a novel method that generates local persistent homology-based data using a modified version of the convolution operator called \textit{Persistent Homology Convolutions}. This method captures information about the locality and translation equivariance of topological features. We perform a comparative study using various representations of histopathology slides and find that models trained with persistent homology convolutions outperform conventionally trained models and are less sensitive to hyperparameters. These results indicate that persistent homology convolutions extract meaningful geometric information from the histopathology slides.
\end{abstract}


\maketitle

\section{Introduction}\label{sec:introduction}

With recent advances in machine learning, there has been a concurrent increase in the availability of large image data sets. Convolutional neural networks (CNNs) are a standard tool for tasks such as image classification, object detection, segmentation, and more. More recently, Vision Transformers (ViTs) have become a popular alternative to CNNs, achieving state of the art results in specific computer vision tasks \citep{wuctvision, wudivision}. The utility  of these models have been seen across numerous domains, especially in the field of computational medicine \citep{farooq2017deep, bullock2019xnet,dozen2020image}. One such example is their use in both labeling and diagnosing diseases such as cancer \citep{ jinnai2020development}. 

\begin{figure}
    \centering
    \includegraphics[width=1.75in]{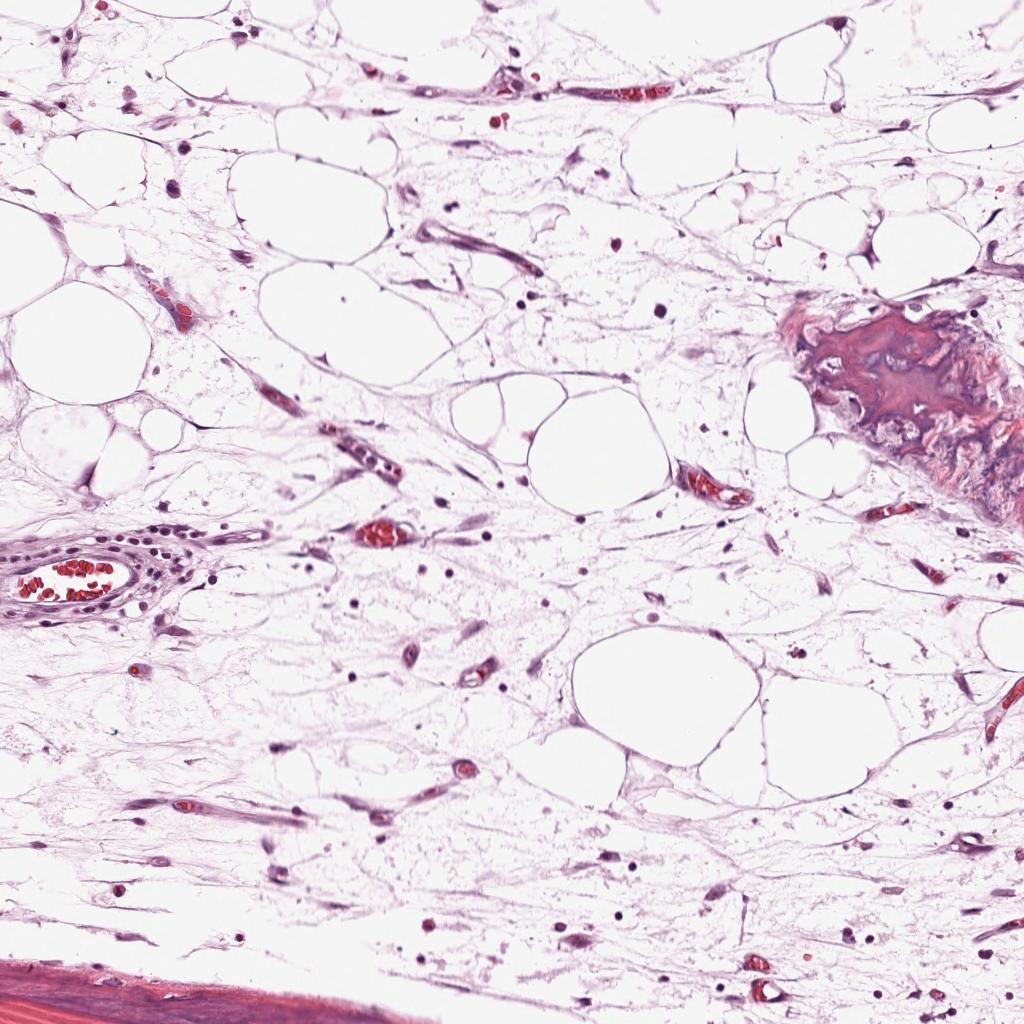}
    \includegraphics[width=1.75in]{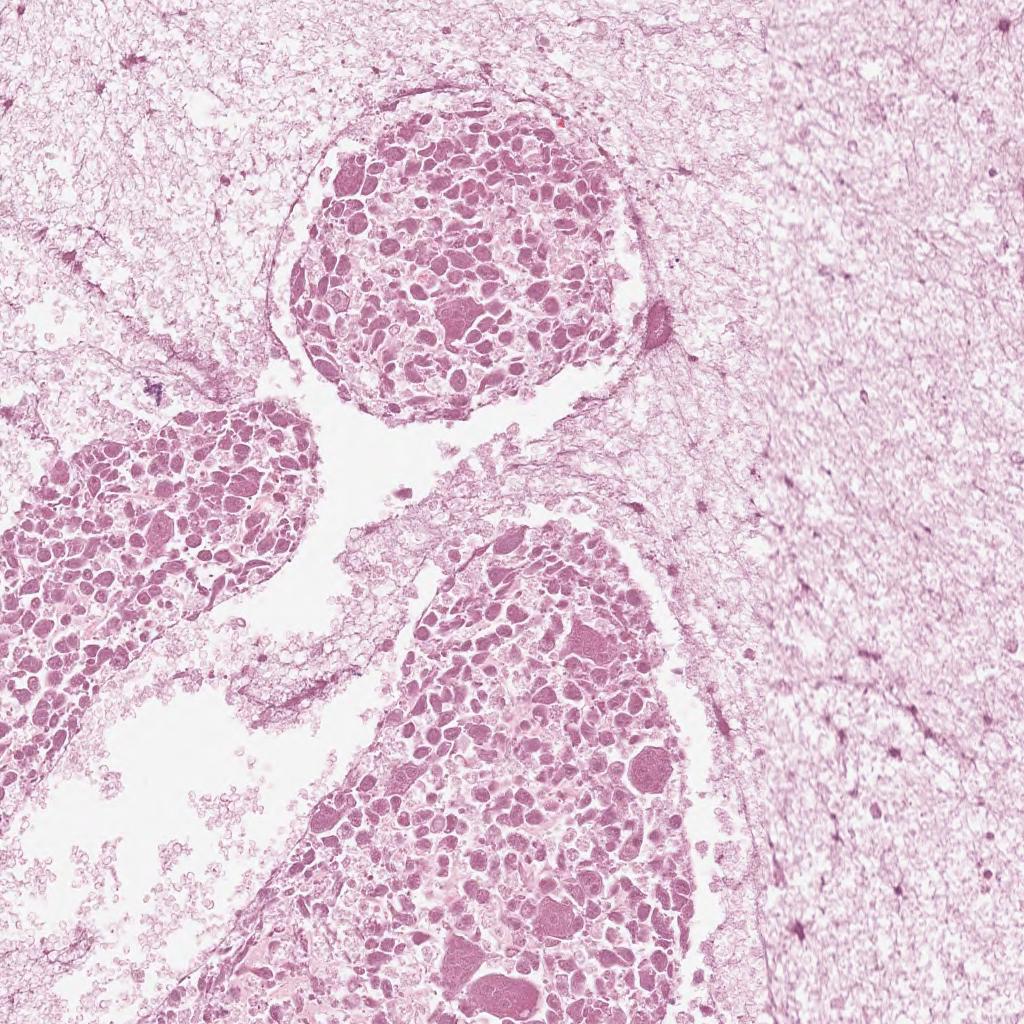}
    \includegraphics[width=1.75in]{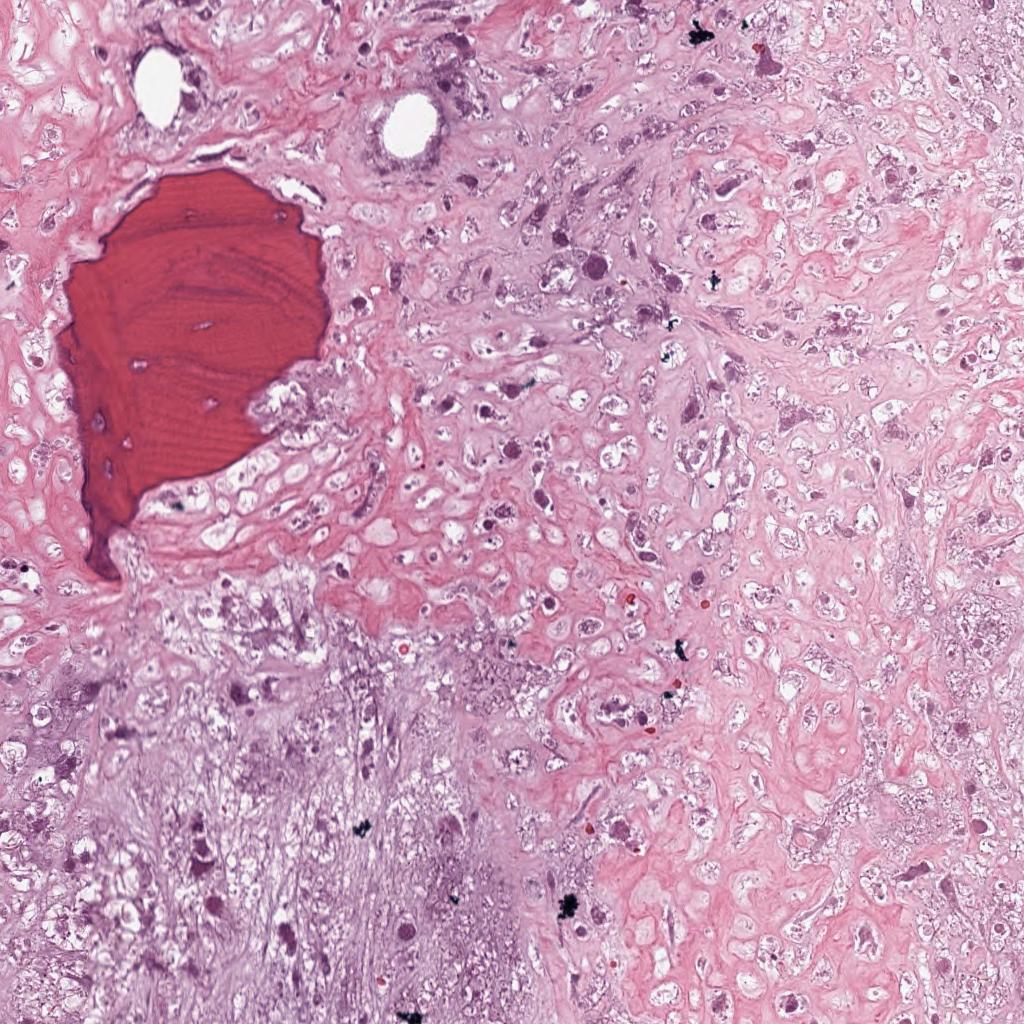}
    \setlength{\belowcaptionskip}{-5pt}
    \caption{\label{fig:pathology} An example of differing tissue structure of between Non-Tumor (left), Necrotic Tumor (center), and Viable Tumor (right) samples in the Osteosarcoma image dataset from \citep{Leavey2019-js,arunachalam2019viable}.}
\end{figure}

Both CNNs and ViTs have been trained to achieve a high degree of accuracy comparable to the performance of a trained pathologist \citep{arunachalam2019viable, komura2018machine, melanthota2022deep,mall2023comprehensive}. However, the CNN pooling operator may alter the relevant geometry within an image \citep{Kepp2019TopologyPreservingSR}. Similarly, the subdivision of images into patches for ViTs alters the geometric structure relative to each patch and may impact performance---for example, by subdividing a region with tumor growth. The geometric structure of cell tissue is an important characteristic in the field of histopathology: a branch of pathology that employs microscopy to examine tissue samples for disease-indicating abnormalities. In diseases like cancer, the cellular abnormalities are often geometric in nature with tissue samples showing varied cell and nucleus size, multinuclation, and a disorganization of tissue structure as illustrated in Figure \ref{fig:pathology}. 

It is natural to ask whether model performance can be improved by reincorporating a summary of the geometry. The data should reflect key information about the architecture of the tissue and shape of the cells. Geometric summaries can be created by utilizing tools from a branch of mathematics known as applied and computational topology; topology is the study of how geometric properties of a space change under continuous transformations. One focus of that field is the development of methods to perform shape analysis (e.g., determine the number of connected components, holes, cavities, etc) \citep{edelsbrunner2022computational}. In recent decades, topology has emerged as an effective tool in data science and deep learning \citep{carlsson2009topology}. In particular, Persistent Homology (PH) can be used to represent geometric features of low-dimensional data and detect the topology of high-dimensional data sets. PH was discovered independently by \citep{frosini1999size}, \citep{robins1999towards}, and  \citep{edelsbrunner2002topological, zomorodian2004computing, edelsbrunner2008persistent}. One appeal of PH is that it provides a vectorizable representation of the shape of data; a perspective that is relevant to histopathology as demonstrated by \citep{lawson2019persistent}. In that specific study, the authors use the persistent homology of a sublevel filtration to quantify the cellular architecture of prostate cancer to accurately predict the Gleason score. Numerous studies have shown utility of PH in histopathology classification \citep{abousamra2023topology, chittajallu2018vectorized, vipond2021multiparameter, lawson2019persistent}. 

\begin{figure}
    \centering
    \includegraphics[width=2.7in]{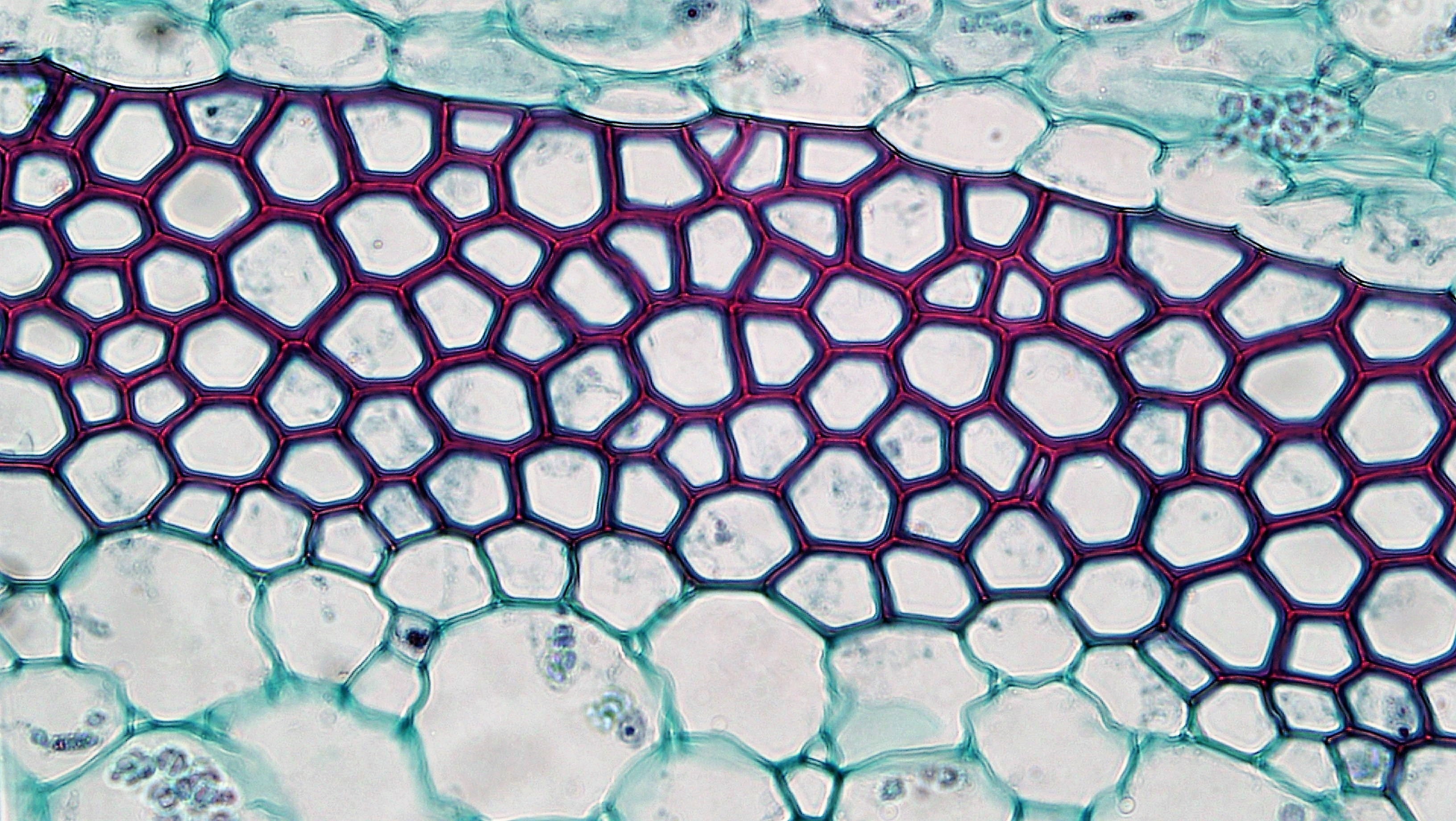}
    \includegraphics[width=2.7in]{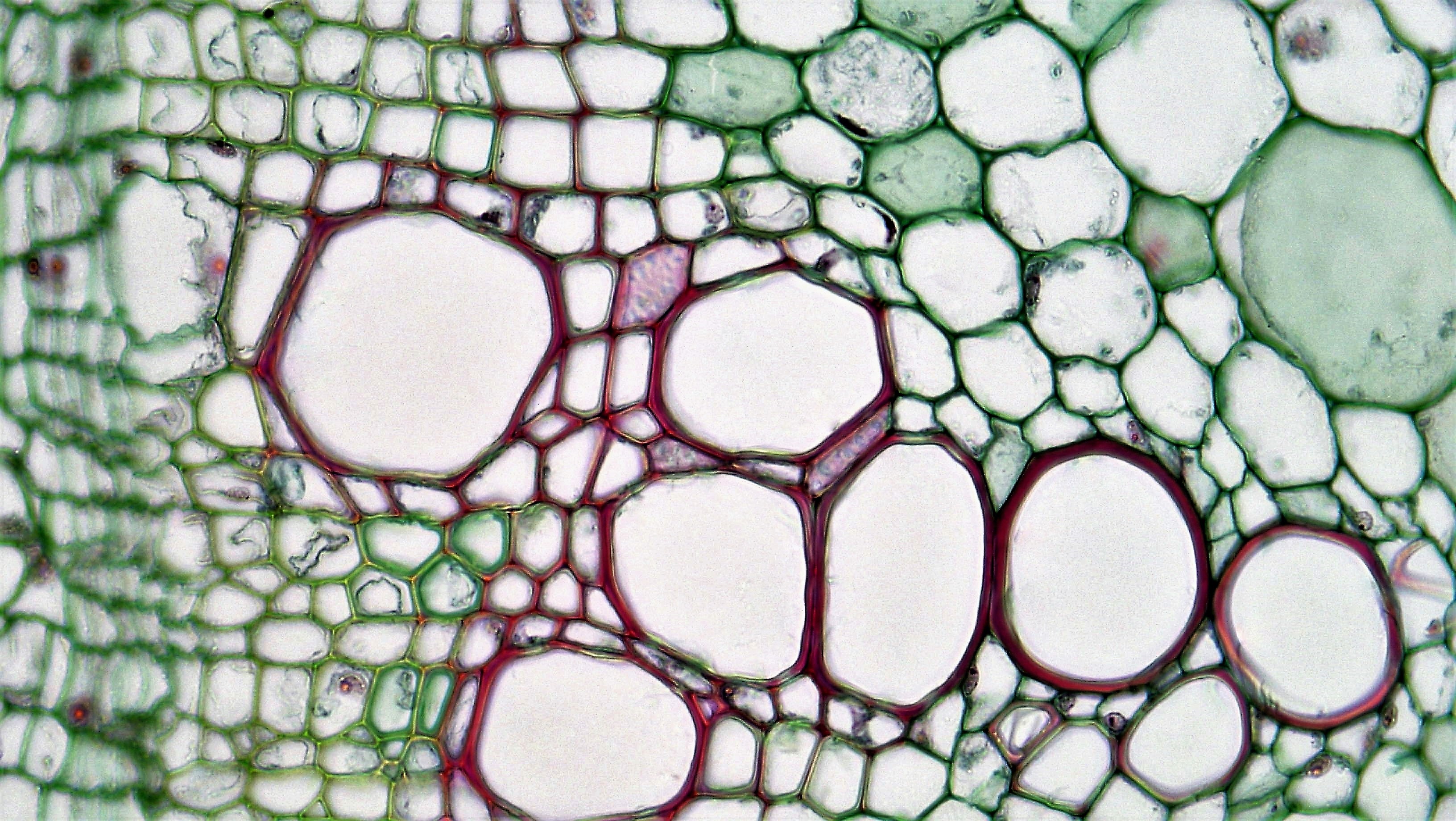}
    \setlength{\belowcaptionskip}{-5pt}
    \caption{\label{fig:local_cell} Images showing the cellular structure of nonwoody stems in a squash (left, Cucurbita sp) and in a castor oil plant (right, Ricinus communis). The tissues contain similar numbers of smaller and larger cells in noticeably different arrangements. Images were taken from \citep{BCCBioscienceImageLibrary}.}
    
\end{figure}

\citep{QAISER20191} demonstrated that CNN model performance in image classification tasks can be improved by combining PH data with image data, as opposed to training a model with image data alone. In their study, PH was computed ``globally'' in the sense that it was measured at the level of an entire image rather than for smaller image patches. We propose that computing PH ``locally'' --- that is, separately in multiple smaller image patches --- may result in better performance by enabling the retention of information related to the relative placement of topological features and reduce computational run time. We hypothesize that the local arrangement of topological
features is an important characteristic for the differentiation of tissue structure, and that some of this information is not detected by the models. To illustrate this, consider Figure \ref{fig:local_cell}. As we explain below, the one-dimensional persistent homology of each image will contain an interval for each cell summarizing its size and shape. As the two images contain about the same numbers of large and small cells, their global PH will be quite similar. However, their local structure is distinct.

A simple approach to computing local persistent homology data is to subdivide the images into fixed patches as done in a ViT and to perform PH computations on separately these subimages before inputting them into other machine learning pipelines for training. Since the ViT's model architecture includes a self-attention mechanism  this method will accurately describe the local topological features in the image data and their relationships to one another \citep{dosovitskiy2021an}.  However, if a region with tumor growth is at the boundary of a patch, this subdivision may result in the loss of important geometric information. Instead, we propose to compute local persistence over a family of possible overlapping patches, similar to how stride is used with the convolution operator in a CNN. That is, the \textit{Persistent Homology Convolution} (PHC) of an $N \times N$ array localized to an $M \times M$ subwindows of $X$ can be expressed as \[[\mathbf{K} \star_{PH} X]=\sum_{i}{\sum_{j}{ \mathbf{K}\star}\mathbf{vec}(\mathrm{PH}_p(\mathcal{F}(T_{(c \cdot i,c \cdot j)}(X)))}.\]
Here, $c$ is the stride length, $T_{(c \cdot i,c \cdot j)}$ is the restriction of $X$ to an $M \times M$ subimage whose bottom left coordinate is $(c\cdot i, c\cdot j)$, $\mathcal{F}$ is a function assigning a filtration to an $M\times M$ subimage, $\mathbf{K}$ is a kernel matrix, and $\mathbf{vec}$ is a vectorization operator. More detail regarding this equation is provided in Sections \ref{sec:background} and \ref{sec:pc}. In short, PHC provide a summary of the geometry in an image that captures information about its locality and translation equivariance.

To illustrate the effectiveness of PHCs we conduct a comparative study of the performance of CNNs trained using multiple different image data representations. The models are trained on an osteosarcoma dataset to distinguish between Non-Tumor, Non-Viable Tumor, and Viable Tumor classes \citep{Leavey2019-js, arunachalam2019viable}. We perform over 10,000 experiments varying image representations and hyperparameters. We find that models trained on PHCs consistently outperform those trained using other data representations on all metrics. Our best performing model yielded $93.8\%$ accuracy in slide-based classification compared to $91.2\%$ accuracy in a previous study on the same data set applying conventionally trained CNNs with a similar architecture\footnote{They did achieve $93.3\%$ after increasing the sample size by subdividing the images; we have not measured the effect of this on the accuracy of our models.}. The same study also tried multiple other conventional machine learning techniques in conjunction with shape analysis resulting in accuracies between 80.2\% to 89.9\% ~\citep{arunachalam2019viable}.  Our results suggest that persistent homology convolutions reduce the complexity of image data while retaining relevant information about the geometry of histopathology slides. The main contributions to the paper are as follows:

\begin{itemize}
    \item[1.] A mathematical definition of PHC.
    \item[2.] A comprehensive empirical study of models trained with PHC for multi-class histopathological classification.  
    \item[3.] A publicly available repository containing the PHC implementation and experimental setup. 
    \footnote{GitHub: https://github.com/Shrunalp/PHC.git} 
\end{itemize}

\section{Background}\label{sec:background}

\subsection{Persistent Homology of Histopathology Slides}

To summarize the geometry of a cell arrangement in a slide, we apply a hybrid approach that measures the geometric properties of topological features using persistent homology (PH). PH encodes both qualitative (e.g. connectedness of tissue and number of cells) and quantitative  (e.g. height, width, area of cells, color) information about this geometry. It is computed by associating to a data set a sequence of spaces (called a filtration) and measuring how the topology changes through the sequence. There are several different filtrations that can be associated to a data set. 

\begin{figure}
    \centering
    \includegraphics[width=13.5cm]{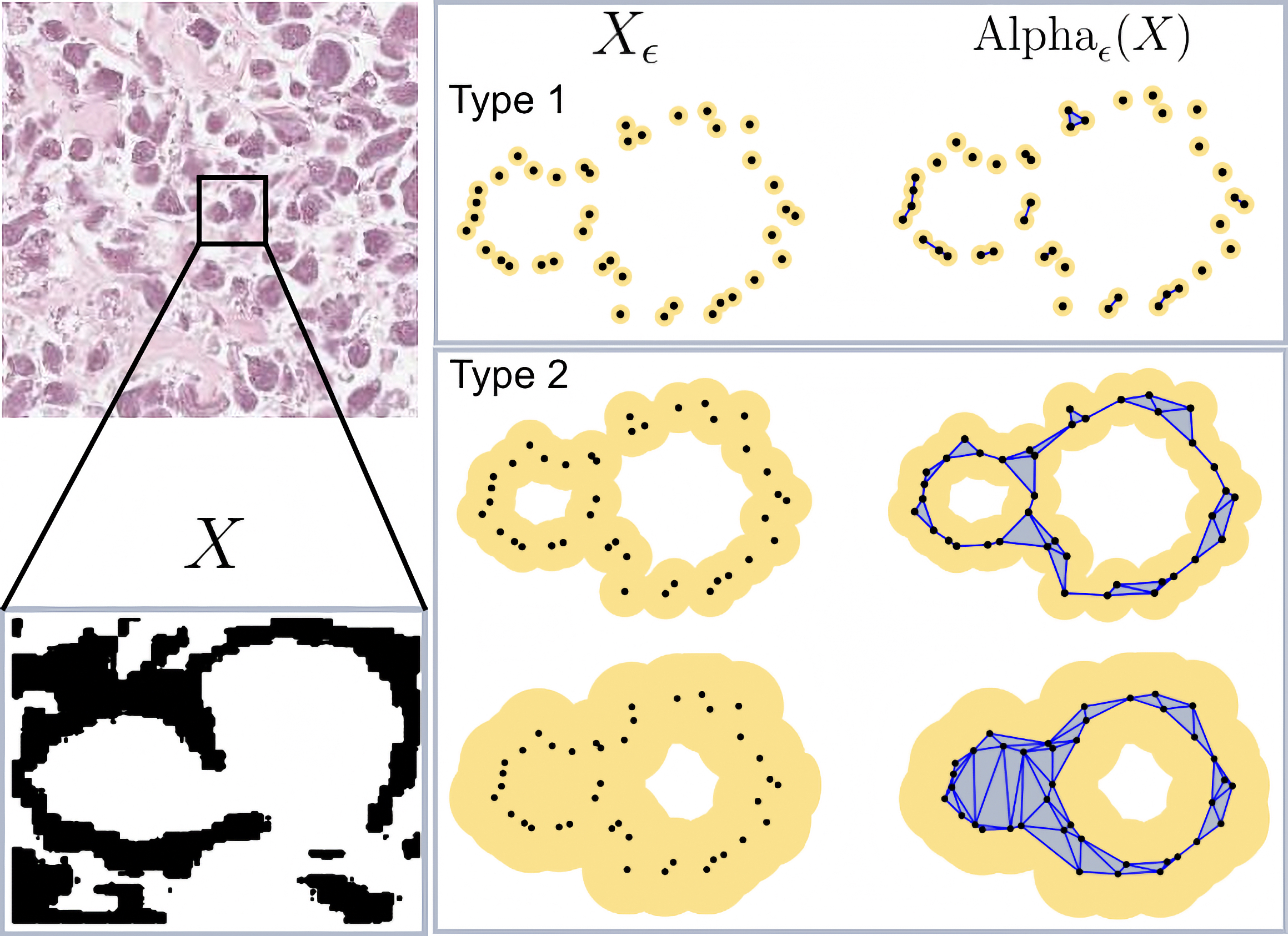}

    \setlength{\belowcaptionskip}{-7.5pt}
    \caption{\label{fig:alpha} Top left: a subimage of a necrotic tumor slide. Bottom left: the threshold set $X$ from the lighter regions around the cells. Right: a simplified representation of $X$; $\mathrm{TPH}_p(X)$ is approximated by taking the alpha complex of $X.$ A visualization of the topological features that emerge is shown for the thickening filtration on the sample (left) and the topologically equivalent alpha complex (right). Type 1 features emerge when connected components merge as $\epsilon$ increases. Type 2 features are born when holes form and die in the union of balls centered at the finite point sample.}
\end{figure}

\subsubsection{The Thickening Filtration and the Alpha Complex}
Most classically, one can associate to a subset $X$ of the plane a sequence of growing neighborhoods $X_{\epsilon}$ consisting of all points within distance $\epsilon$ of $X.$ The persistent homology of $X$ will then measure how components of $X$ merge and how holes in $X$ form and are filled in as $\epsilon$ increases. In this context, persistent homology measures the \textit{birth} and \textit{death} of topological features. For point cloud data in low-dimensional Euclidean space, this can be computed in practice using the alpha complex \citep{edelsbrunner1994three}. 

To apply this to a grayscale image, the image is thresholded to yield a subset $X$ of the unit square $\brac{0,1}.$ $X$ consists of a union of pixels. In practice, the persistent homology of the alpha complex of the set of centers of these pixels yields a collection of intervals that is, in a well-defined sense~\citep{cohen2005stability}, very close to the persistent homology of the thickening filtration $\set{X_{\epsilon\geq 0}}$ of $X.$ We provide an informal, geometric description of this persistent homology specialized to this setting.

\begin{proposition}
\label{prop:feature_typealpha}
The Persistent Homology Diagram of the thickening filtration of $X\subset \brac{0,1}^2$ is a multiset whose elements consist of intervals of the form $(b,d)$. For each interval, the initial coordinate $b$ represents the value of $\epsilon$ at which a feature is born and the second coordinate $d$ is when it dies. There are two distinct types of features measured by $\mathrm{TPH}_p(X)$. The former is of dimension $p=0$ and the latter is of dimension $p=1$:

\begin{itemize}
    \item[Type 1.] An interval $(0, d) \in \mathrm{TPH}_p(X)$ for each component of $X.$ The component merges with another in the thickened set $X_{d}.$ A single component of $X$ lasts throughout the filtration and has $d=\infty.$ 
    \item[Type 2.] An interval $(b,d)\in \mathrm{TPH}_p(X)$ for each bounded component of the complement that appears in the thickened sets $X_{\epsilon}.$ The bounded components of $\R^2\setminus X$ correspond to intervals with $b=0.$ An birth time $b$ of an interval $\paren{b,d}$ with $b>0$ corresponds to the time when a ``gulf'' of $X$ closes to form a component of $\R^2\setminus X_{b}.$ The component disappears in the thickened set $X_d.$ Roughly speaking, the death time $d$ measures the largest enclosed ball that can fit inside the gulf or hole of $X$ corresponding to the feature.  
\end{itemize}
\end{proposition}

In our study, $X$ is the subset of pixels of a histopathology slide whose grayscale values are lighter than a fixed threshold. For example, the lighter regions in the right most slide of Figure \ref{fig:pathology} correspond to the space between cells, isolated by surrounding tissue matter. As such, Type 1 features measure the density of cells and the uniformity of their arrangement. Type 2 features correspond to cells or subcellular structures that are darker than the neighboring matter (multinucleation); the birth and death values measure information about the size and shape of those features. As we describe above, we approximate $\mathrm{TPH}_p(X)$ using the alpha complex of a finite subset $\set{x_1,\ldots,x_n}$ of $X$; the persistent homology of the alpha complex is the same as $\mathrm{TPH}_p(\set{x_1,\ldots,x_n}).$ See Figure \ref{fig:alpha}.

\subsubsection{The Extended Lower Star Filtration}
Alternatively, the lower star filtration of a grayscale image $X$--- denoted $\mathrm{St}_{\underline{\hspace{2mm}}}(X)$ ---is the sequence of sets $X_{\rho}$ consisting of the union of all pixels whose grayscale value is less than or equal to $\rho$ (ranging from black pixels assigned $0$ to white pixels assigned $1.$). That is, we associate to the image $X$ the filtration of level sets $X_{\rho}=\{x\in X \colon f(x)\leq \rho\}.$ In the bottom two rows of Figure~\ref{fig:persistence_ex}, this is shown by progressively increasing the sublevels to observe when individual features (new components or holes) are \textit{born} and when they merge with other feature or \textit{die}. This filtration was applied to a classification task on prostate cancer histology slides in \citep{lawson2019persistent}. Additional features are captured by the \textit{extended persistence} of this filtration, which scans ``back down'' through the superlevel sets $X^{\rho}$ the superlevel set $\{x\in X \colon f(x)\geq \rho\}$ of $f.$ More formally, the extended lower star filtration is defined by appending the quotient spaces $\paren{\brac{0,1}^2/X^{\rho}:1\geq \rho \geq 0}$ (in descending order) to the collection of sublevel sets, where $\brac{0,1}^2/X^{\rho}$ is obtained from the square $\brac{0,1}^2$ by collapsing the superlevel set $X^{\rho}$ to a point. See Figure~\ref{fig:xph}.

We provide an informal, geometric description of the extended persistence diagram  $\mathrm{XPH}_p(f).$ We specialize to the case where the domain of $f$ is the unit square; this avoids complications and allows for an account  which does not not require knowledge of algebraic topology.  The technical background is covered extensively elsewhere; see \citep{agarwal2004extreme, edelsbrunner2022computational, cohen2009extending, turner2024extended} for a rigorous treatment of the extended persistent homology of a function and \citep{robins2011theory, delgado2014skeletonization} for accounts specific to the sublevel set filtration of a grayscale image. 

Consider an $N\times N$ grayscale function, and subdivide $\brac{0,N}^2$ into $N^2$ unit squares. The image induces a step function $f:\brac{0,N}^2\to\brac{0,1}$ that assigns to each unit square the corresponding grayscale value. The Extended Persistent Homology of $f,$ denoted $\mathrm{XPH}_p(f)$, measures how the topology of sublevel sets $f^{-1}\paren{\brac{0,l}}$ and superlevel sets $f^{-1}\paren{\brac{s,1}}$ change with the parameters $l$ and $s.$ To give an elementary description, we replace $f$ with a smooth function $\tilde{f}$ whose persistent homology is infinitesimally close to that of $f.$ First, $f$ is not continuous so we linearly interpolate between different values of $f$ in an infinitesimal boundary of each unit square to yield a continuous function. Next, we round off the corners of the square $\brac{0,1}^2$ to yield a domain $D$ with a smooth boundary. Finally, we infinitesimally perturb the function to obtain a smooth function $\tilde{f}:D^2\to\brac{0,1}$ with no degenerate critical points (that is, so that the Hessian matrix is nonsingular at each interior critical point, and the single variable function obtained by restricting $\tilde{f}$ to the boundary has no inflection points). $\tilde{f}$ is called a \emph{Morse function}. By applying an additional perturbation if necessary we may assume that $\tilde{f}$ takes distinct values at each critical point. 

\subsubsection{Ordinary Persistent Homology} 
\begin{figure*}
    \centering
    \includegraphics[width=13.5cm]{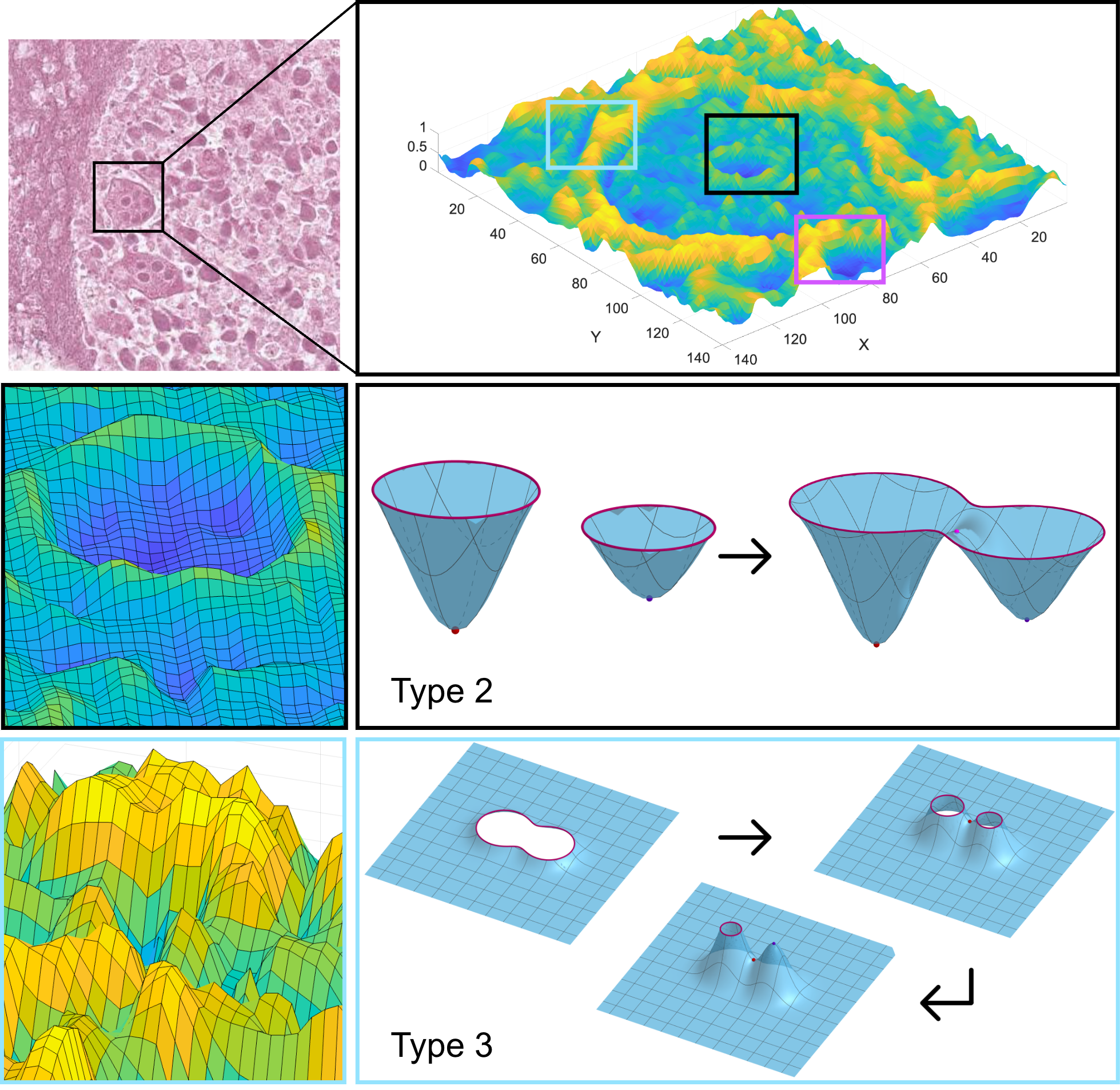}

    \setlength{\belowcaptionskip}{-7.5pt}
    \caption{\label{fig:persistence_ex} Top left: histopathology slide of necrotic tumor cells. Top right: the graph of $\tilde{f}$ in the vicinity of a cell with multi-nucleation. $\tilde{f}$ possesses many minima, saddle, and maxima; pairs of neighboring critical points correspond to different types of persistent homology intervals. Each colored box on the density plot has a simplified representation of the geometric feature below that appears in the filtration. There is a single Type 1 feature (magenta box) corresponding to the absolute minimum of $\tilde{f}.$ Type 2 features (black box) pair local minima with neighboring saddle points (or maxima on the boundary); they represent cellular or subcellular structures darker than neighboring pixels. The illustrated  feature quantifies multi-nucleation. Type 3 features (blue box) pair local maxima with neighboring saddle points; these pairs capture measurements related to structures lighter than their neighbors. Here, a void between cells gives rise to a Type 3 feature.}
\end{figure*}

We begin by describing the ordinary persistent homology of $\tilde{f}$ in terms of pairs of critical points. To aid visualization, it is useful to picture the sublevel sets $D_l$ as not living in the planar set $D$ but rather in the graph of $\tilde{f}$ in three dimensional space, as shown at the top right of Figure~\ref{fig:persistence_ex}. As such, we sometimes refer to $\tilde{f}\paren{x}$ as the ``height'' of $\tilde{f}$ at the point $x.$

First, the $0$-dimensional persistent homology measures how components of the sublevel sets $D_l\coloneqq \tilde{f}^{-1}\paren{\brac{0,l}}$ appear and merge as $l$ increases from zero to one. A new component emerges at each local minimum of $\tilde{f},$ whether they are on the boundary or the interior. The component born at the absolute minimum $b_0$ of $\tilde{f}$ never disappears, so it corresponds to an an interval $\paren{b_0,\infty}$ in dimension $p=0.$  Other local minima are paired either with saddle points of $\tilde{f},$ as illustrated in the second row of Figure~\ref{fig:persistence_ex}, or with local maxima on the boundary. This can be seen as follows.  Suppose $x_0$ and $x_1$ are two local minima of $\tilde{f}.$ Since $D_1$ is connected, there is an $0<l'<1$ so that $x_0$ and $x_1$ are in different connected components of $D_l$ for $l<l'$ but in the same component of $D_{l'}.$ A path from $x_0$ to $x_1$ must pass through a point $x_2$ so that $\tilde{f}\paren{x_2}=l'.$ If $x_2$ is in the interior of $D$ then there must be directions in which $\tilde{f}$ is both increasing and decreasing at $x_2$ (if not, $x_0$ and $x_1$ would be connected in a lower level set) so $x_2$ must be a saddle point of $\tilde{f}.$ Otherwise, if it is on the boundary it must be a local maximum.

Now, consider a local maximum $x$ of $\tilde{f}$ in the interior of its domain, such as the one illustrated in  the bottom row of Figure~\ref{fig:persistence_ex}. The sublevel set $D_{f\paren{x}-\epsilon}$ contains a loop around $x$ which cannot be contracted to a point (otherwise, $\tilde{f}$ would not be a local maximum). This loop generates a one-dimensional homology class which ``dies'' at $d:=\tilde{f}\paren{x}.$ $D_0$ does not contain any cycles, so this one must be ``born'' at some point. If we replaced $\tilde{f}$ with $-\tilde{f}$ the local minimum $x_0$ of $-\tilde{f}$ would be paired with a saddle point (or boundary local maximum) of $-\tilde{f}$ at some point $x_1.$ This is also a saddle point (or boundary local minumum) of $\tilde{f}$ and it turns out the the cycle which dies at $d$ first appeared at $b:= \tilde{f}\paren{x_1},$ perhaps splitting from another loop. 

\begin{proposition}
\label{prop:feature_type}
Let $\tilde{f}:D\to\brac{0,1}$ be as above. The $p$-dimensional Persistent Homology Diagram, denoted $\mathrm{PH}_p(\tilde{f})$, is a multiset whose elements consist of intervals of the form $(b,d)$. For each interval, the initial coordinate $b$ represents the height that a feature is born and the second coordinate $d$ is when it dies. There are three distinct types of features measured by $\mathrm{PH}_p(\tilde{f})$. The first two are of dimension $p=0$ and the last is $p=1$:

\begin{itemize}
    \item[Type 1.] A single interval $(\min\{\tilde{f}(x)\}, \infty) \in \mathrm{PH}_0(\tilde{f})$ for the single component of $D.$
    \item[Type 2.] An interval $(b,d)\in \mathrm{PH}_0(\tilde{f})$ for each local minimum $x_0$ of $\tilde{f}$ other than the global minimum. The feature is born at height $b=\tilde{f}\paren{x_0}$ and dies at the height $d$ of a saddle point or a boundary local maximum where the component merges with another one. 
    \item[Type 3.] An interval $\paren{b,d}\in \mathrm{PH}_1(\tilde{f})$ for each local maximum of $\tilde{f}$ in the interior of its domain. The feature is born at the height $b$ of a saddle point or boundary local minimum and dies at the height $d$ of the local maximum.
\end{itemize}
\end{proposition}

Larger values of $p$ correspond to higher-dimensional topological features which do not occur in this setting. Each type of feature captured Proposition \ref{prop:feature_type} has the following biological interpretations. 

Type 2 features correspond to various cellular and subcellular structures that are darker than their neighbors; the birth and death values respectively encode information about the tissue type of the structure and its neighbors. Type 2 features for the osteosarcoma dataset indicate the presence of a cell, nuclei, nuclear fragmentation (karyorrhexis), or other distinct tissue matter. The darkest structures in this histopathology data set are nuclei, and multinucleation events are recorded by Type 2 features with low death times (as shown in the second row of Figure \ref{fig:persistence_ex}). Type 3 features correspond analogously to structures that are lighter than their neighbors, such as empty voids between cells as illustrated in the third row of Figure \ref{fig:persistence_ex}). Intervals of both types of features whose corresponding critical points are in the interior of $D$ represent structures contained in the interior of $D,$ whereas the others encode information about cellular and subcellular structures adjacent to the boundary.

\subsubsection{Extended Persistent Homology} The Extended Persistent Homology  $\mathrm{XPH}_p(\tilde{f})$ captures additional information about the geometry of the the level sets of $\tilde{f}.$ Recall that $D^l=f^{-1}\paren{\brac{l,1}}$ and that the quotient space $D/D^{l}$ is obtained by collapsing $D^l$ to a single point. The filtration proceeds to collapsing more and more of $D$ to a point. A topological feature which appears in $D/D^{b}$ and disappears in $D/D^{d}$ is assigned an interval $\paren{b,d};$ note that these intervals always have $d<b$ whereas ordinary persistent homology intervals satisfy $b<d.$ 

\begin{figure}
    \centering
    \includegraphics[width=13.5cm]{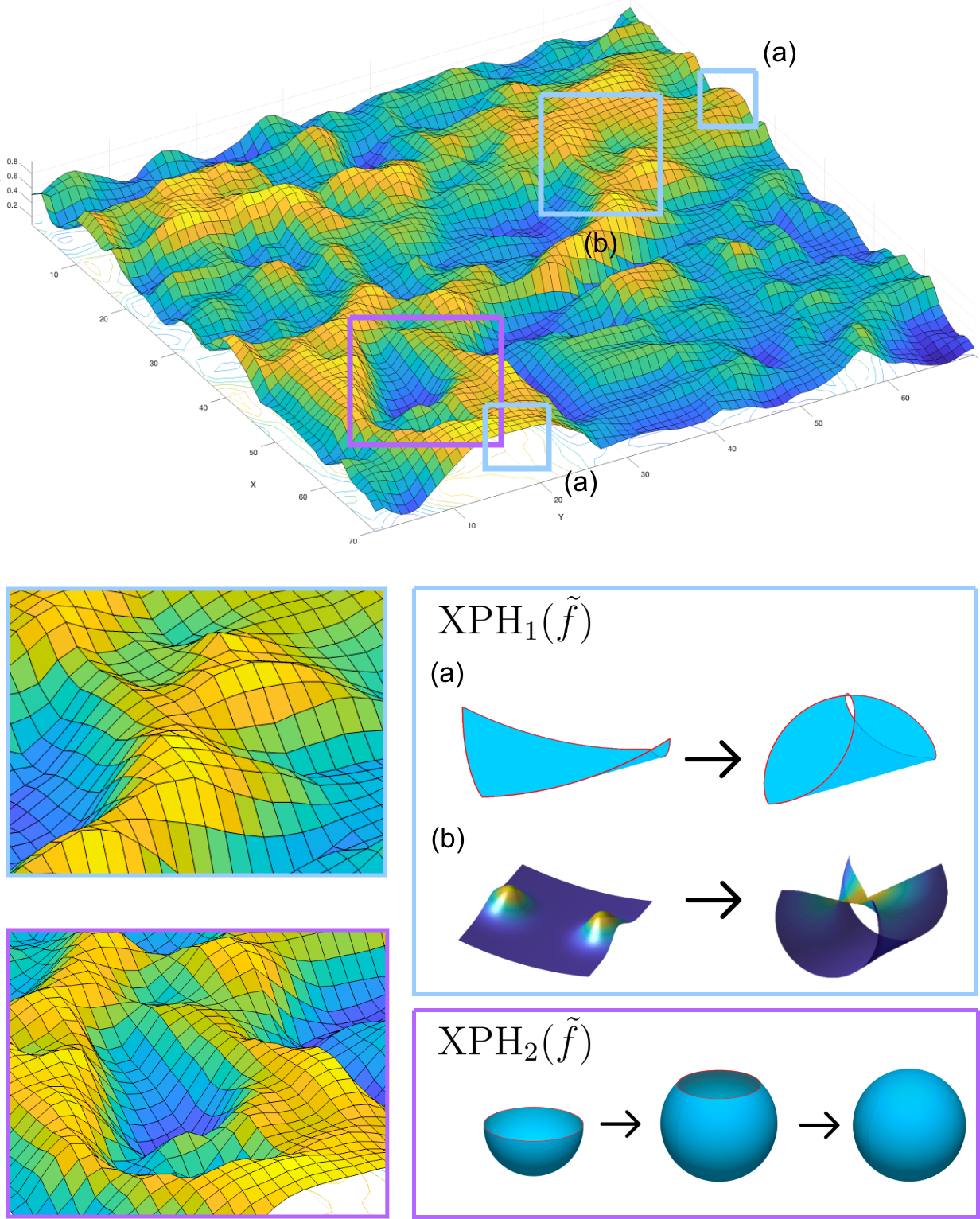}

    \setlength{\belowcaptionskip}{-7.5pt}
    \caption{\label{fig:xph} Top: density plot of a necrotic histopathology slide. Bottom: visual representation of features generated by quotienting the superlevel sets. There are two instances in which loops are generated in $\mathrm{XPH}_1(\tilde{f})$ as shown in the blue box. This occurs when the maxima on the boundary (shown by (a)) or the relative maxima within the surface (shown by (b)) are identified in the quotient. Lastly, voids are created if there exists a cavity in the surface, allowing for the surrounding ridge to collapse to a point, creating a sphere as shown in the magenta box. These voids correspond with features in $\mathrm{XPH}_2(\tilde{f})$. }
\end{figure}

Two additional types of topological changes occur in the extended filtration. First, suppose that $f$ achieves its global maximum at $x_0$ and a local maximum at a point $x_1.$ $x_0$ and $x_1$ are connected in $D$ but not in the superlevel set $D^{f\paren{x_1}}.$ Collapsing  $D^{f\paren{x_1}}$ to a point forms a loop in the quotient space $D/D^{f\paren{x_1}}.$ This is a one-cycle in the extended persistent homology which persists until it is collapsed to a point, either at a saddle point or a boundary local minimum. If the feature is born at an interior global maximum at height $b=f\paren{x_1}$ and dies at a saddle point at height $d<b,$ the extended persistence interval $\paren{b,d}\in \mathrm{XPH}_1(\tilde{f})$ is paired with a corresponding interval $\paren{d,b}\in  \mathrm{PH}_1(\tilde{f}).$ See the first two rows of Figure~\ref{fig:xph}.

Let $x_1$ be a local minimum of $\tilde{f}$ in the interior of $D$ and set $d=\tilde{f}\paren{x_1}.$ For small $\epsilon,$ the superlevel set $D^{d+\epsilon}$ separates $x_1$ from the other local minima of $\tilde{f}.$ Thus, collapsing $D^{d+\epsilon}$ to a point creates ``pocket'' around $x_1$ in $D/D^{d+\epsilon}.$ This is a two-dimensional topological feature. If we look for the first ``neck'' collapsed to form this void in  $D/D^{b}$ we see that $\tilde{f}$ has either a saddle point or boundary local maximum at height $b.$ In the former case, this interval is paired with a corresponding one $\paren{d,b}\in \mathrm{PH}_2(\tilde{f}).$ A topological feature of this type is illustrated in the third row of Figure~\ref{fig:xph}. 

\begin{proposition}
\label{prop:extended_feature_type}
The $p$-dimensional Persistent Homology Diagram $\mathrm{XPH}_p(\tilde{f})$ contains the intervals of $\mathrm{PH}_p(\tilde{f}),$ and two additional types of intervals. They are of dimensions $1$ and $2,$ respectively.

\begin{itemize}
    \item[Type 4.] An interval $(b,d)\in \mathrm{XPH}_1(\tilde{f})$ for each local maximum $x_0$ other than the global maximum. The feature is born at height $b=\tilde{f}\paren{x_0}$ and dies at the height $d$ of a saddle point or a boundary local minimum. 
    \item[Type 5.] An interval $\paren{b,d}\in \mathrm{XPH}_2(\tilde{f})$ for each local minimum of $\tilde{f}$ in the interior of its domain. The feature is born at the height $b$ of a saddle point or boundary local maximum and dies at the height $d$ of the local minimum.
\end{itemize}
\end{proposition}

The biological interpretation of the additional intervals of $\mathrm{XPH}_p(\tilde{f})$ is similar to that given for $\mathrm{PH}_p(\tilde{f})$ at the beginning of the previous section. It captures the same information about structures in the interior of $D,$ and different information about structures adjacent to the boundary.

\subsection{Vectorization of Persistence Data} 

As described above, the PHC data consists of collections of sets of intervals, one for each window. This incompatible with the machine learning algorithms we intend to use, which take vectors as input. In particular, note that the number of intervals may vary between windows. As such, we convert the PDs into vectors in a fixed-dimensional Euclidean space. There are multiple vectorization methods for persistence diagrams \citep{ali2023survey}. A precise mathematical definition of the vectorization operation is found in \citep{carriere2020perslay}.

There are multiple vectorization methods for persistence diagrams. In our study, we utilize the notion of a \textit{persistence image} developed by \citep{adams2017persistence}.  This representation converts a diagram into a finite-dimensional vector in $\R^{n\times n}$ --- denoted   $\mathbf{I}_{n \times n}(z)$ --- that may be interpreted as a discretized heat map of the scatter plot of interval endpoints of $\text{XPH}_1(f)$. See~\citep{adams2017persistence} for a complete definition. The persistence image is stable with respect to input noise. In one study, it was shown to perform better than alternative vectorizations in machine learning applications \citep{adams2017persistence}.

\section{Persistent Homology Convolutions} \label{sec:pc}

The convolution operator in CNNs is a method to synthesize and extract relevant attributes in an image for training. Suppose that $X$ is an $N \times N$ grayscale image, that is, its pixel value at coordinates $(x,y)$ are $X(x,y) \in [0,1]$ for $0\leq x,y \leq N$ and zero otherwise. Given a kernel $k(x,y)$ the (discretized) convolution operator used in a CNN can be expressed as \begin{equation}
    \label{eqn:conv}[k \star X](x,y)=\sum_{\tau_x \in \Z}{\sum_{\tau_y \in \Z}{ k(x,y)X(x-\tau_x,y-\tau_y)}}.
\end{equation} Intuitively, the kernel $k(x,y)$ is used to extract relevant features from $f(x,y)$ by performing computations with the kernel as it translates over the entire image. The output of this operation is a feature map that contains synthesized information regarding the original image and is then subsampled (pooled) and is then (generally) repeated multiple times before being feed to the multilayer perception.  The convolution operator is translation invariant but the process of convolving and pooling does not necessarily preserve the topology in an image. To address these shortcomings, we present \textit{Persistent Homology Convolutions}. This operation shares similar attributes to the classic convolution operator, but instead measures local topological features while maintaining that the operator is translation equivariant to such measurements. We present this idea on an $N \times N$ grayscale image $X \in \mathcal{G}_N$ where $ \mathcal{G}_N$ denotes the set $[0,1]^{[[N]] \times [[N]]}$ with $[[N]]=\{0,1,\dots, N\}$.

   \begin{definition}

    Let $X \in \mathcal{G}_N$ be an $N \times N$ grayscale image. For $M<N$ consider the function $T_{(x,y)}\colon \mathcal{G}_N \to \mathcal{G}_M$ which maps $X$ to an $M \times M$ grayscaled image by translating the  $(x,y)$ coordinate of $X$ to the origin and restricting computations to $[[M]]^2$. Given a kernel matrix $\mathbf{K}$ and a function $\mathcal{F}$ assigning a filtration to an $M\times M$ subimage, we define the \textbf{Persistent Homology Convolution} operator as 
    \begin{equation}    
    [\mathbf{K} \star_{PH} X]=\sum_{i = 1}^{K}{\sum_{j =1}^{K}{ \mathbf{K}\star }\mathbf{vec}(\mathrm{PH}_p(\mathcal{F}(T_{(c \cdot i, c \cdot j)}(X)))} \label{eqn:phc}\end{equation} 
    where $K = \lfloor \frac{N-M}{c} \rfloor$ for some $1 \leq c \leq M$ and $\mathbf{vec}$ is an operator that maps persistent diagrams to vectors in Euclidean space.
  \end{definition}

Here, $c$ acts as the stride length used to translate the window. To prevent the window from being translated outside of the boundary of the image, the $x$-coordinate and $y$-coordinate translations are bounded between $1 \leq i,j \leq K$. In our study, the filtration $\mathcal{F}(T_{(c \cdot i,c \cdot j)}(X))$ corresponds to either the alpha complex, or a modified version of the extended lower star function on the $M \times M$ subimage of $X$ (as described in the next section). In our study, we focus on persistence computed in dimension one as it yielded the best results. The localized PH computation is expressed as $Y= \mathbf{I}_{n \times n}(\mathrm{XPH}_1(f(T_{(32 \cdot i,32 \cdot j)}(X))))$ where the stride was set to 32 (see Section \ref{section:implement}). Similarly, the methods used by both \citep{QAISER20191} and \citep{lawson2019persistent} to compute persistence on histology slides can be presented as $\mathrm{PH}_p(\mathrm{St}_{\underline{\hspace{2mm}}}(X))$ where $\mathrm{St}_{\underline{\hspace{2mm}}}(X)$ represents the lower star filtration on the magnitude of pixels in $X$ and $p=0,1$. The map $T_{(i,j)}$ is removed from the expression since the computations are performed globally on $X$ rather than on windows.

Using persistence images, the output local PH computation is a topological feature vector in $Y \in \R^{n\times n}$ (as stated above). The $n \times n$ array is then convolved with the kernel matrix $\mathbf{K}$ whose entries are $\mathbf{K_{ij}}=k(w,T_{\paren{k, \ell}}) = w^T \cdot  T_{\paren{k, \ell}}$ where $w \in \R^{q\times q}$ is a weight matrix and $T_{\paren{k, \ell}}$ is the translation map restricted to a $q \times q$ window. Equation \ref{eqn:conv} is applied to yield the full expression in Equation \ref{eqn:phc}. In CNNs, the weight matrix is implicitly determined and optimized in the kernel matrix $\mathbf{K}$ via backpropagation during model training. The kernel performs feature extraction on the vectorized data to optimally search for the relevant topological signatures which it uses to distinguish between the tumor classes. We hope to use a similar approach in the future to adaptively optimize a weighting function in the vectorization operation.

\section{Experiments}
\label{sec:method}

\subsection{Experimental Setup}
We study three distinct representations of a histopathology slide: a grayscaled, thresholded image, global persistent homology computed for the entire image, and local persistent homology in the form of PHCs. Our objective is to compare the performance of models trained with each of these three representations with respect to multiple performance metrics. We also compare model performance of PHCs based on two different filtrations. 

\subsubsection{Histopathology Dataset}
We apply CNNs to a dataset of histopathology slides for the diagnosis of Osteosarcoma (Ost.), a rare form of bone cancer. It is available at the Cancer Imaging Archive \citep{clark2013cancer,Leavey2019-js}. The dataset consists of 1144 RGB images with resolution (1024, 1024, 3). The images are separated into three classes: non-tumorous (47\%), non-viable (or necrotic) tumor (23\%), and viable tumor (30\%). It is desirable that the dataset is balanced in each class with roughly 381 images (roughly a third of the total number of images). To achieve this, we resample 381 non-tumorous images and perform image augmentation (rotations and reflections) on the remaining classes to balance them. Since homology is invariant to rotations and reflections, the global PH data will contain multiple identical summaries. This illustrates that certain forms of data augmentation cannot be used when persistence is computed globally without introducing redundancies. The resulting experimental dataset contains 1143 RGB images and serves as our base-line comparison against the PH-based data. 

\subsubsection{PHC Data Generation}
\label{sec:PC_data}
 There are several preprocessing steps that are taken to condition the image data before computing persistent homology. These standard methods are applied to enhance the geometric features in the tissue structure, enabling more accurate and robust summaries. Note that we apply the same preprocessing to the data used for each of the three methods (that is: global persistence, local persistence) to ensure a fair comparison. The basic procedure is as follows:

\begin{itemize} 
    \item[1.] Images are grayscaled and resized to minimize topological summary compute time.
    \item[2.] 
    The image is conditioned (truncated/thresholded) and the tissue architecture is emphasized using erosion.
    \item[3.] All remaining pixels (the boundary of the cells) are appended to a data structure known as a simplex tree along with edges from neighboring pixels whose values are nonzero. 
    \item[4.] Persistence is computed on the simplex tree and the summaries are vectorized into a persistence image.
\end{itemize}

\begin{figure*}
\centering
    \includegraphics[width=5.6in]{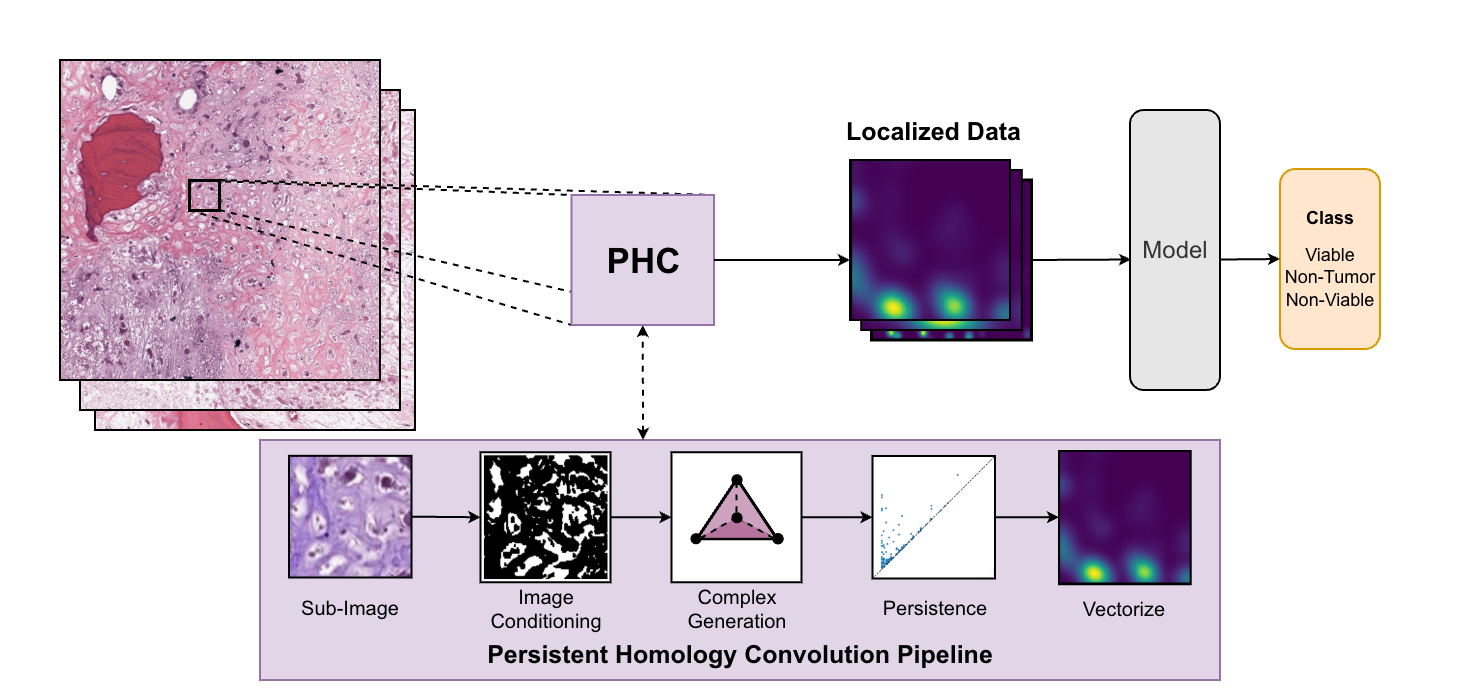}
    \setlength{\belowcaptionskip}{-5pt}
    \caption{Our preprocessing pipeline for image classification using PHC. As the convolution operator slides across the image several sub-tasks are performed. The sub-image is first conditioned before being used for complex creation. Next, the persistence of the complex is computed and vectorized. The data is collated and used for model training.}
    \label{fig:pc_pipeline}
\end{figure*} Multiple color channels, although generally useful for image classifications with CNNs and ViTs, do not contribute any meaningful topological information with our methodology. Additionally, PH computations are computationally expensive. To reduce computation time we resized all grayscale images from $1024\times 1024$ to $512\times 512$ for global and local persistence computations.

The gray scale images are then conditioned used one of two techniques: thresholding or truncating. Thresholding is applied when computing the alpha complex, and truncation is applied for the other filtrations. For truncation, all pixel values above a parameter $\rho$ are retained, including their pixel intensity value. Both methods segment the tissue architecture with the former resulting in a binary image. This removes extraneous features, such as small, noisy holes between cells and within cells. We experimented with multiple truncation/threshold values ranging from 180-200 and found $k=200$ yielded the best performance. We also found that performance was worse if no truncation or thresholding was applied.  

Erosion is applied once to ``thicken" the boundary of the cells and fill in any small, ``noisy" holes using a $2 \times 2$ kernel. 

We compute the persistent homology using multiple filtration methods. Please see Appendix \ref{sec:appendix} for the full list of results. We restrict our focus here to the two best performing filtrations: the alpha complex and a modified version of lower star filtration.  The alpha complex filtration is computed by inputting the centers of the pixels which remain after thresholding into the implementation in GUDHI~\citep{gudhi:urm}. The second is a modified form of extended persistent homology on the lower star filtration. Here, the conditioned image data from Step 2 is converted into a data structure known as simplex tree, a trie data structure used to efficiently represent any general simplicial complex \citep{boissonnat2014simplex}. We create a simplex tree by adding vertices for each pixel, and edges between adjacent pixels in the truncated set. Each vertex is given a filtration value that correspond to the grayscale value of the pixel. For edges, its filtration value is determined by the max of its two vertices. This is called the \textit{adjacency complex} of the image. The extended persistent homology of this filtration yields the same information as the birth times of the persistent homology intervals of the one described in Propositions~\ref{prop:feature_type} and~\ref{prop:extended_feature_type}. We found that this yielded better performance than the extended persistent homology of the lower star filtration. 

Extended persistence is then computed using either the entire simplex tree or locally with PHC. In the case of PHC, the region $U$ was chosen to be a square window of size $32 \times 32$. We fix the stride to $32$ as this yielded the best results. The final persistence data was then vectorized into a persistence image. There are various resolution sizes that could be chosen for persistence image $\mathbf{I}_{n\times n}(z)$. Performance was consistent for any \[ n \in \{10, 15, 20, 25\}\] so $n=20$ was fixed. 

The output PHC data of a single image is a 3-dimensional array of the form $(256, 400, 1)$ which corresponds to $256$ persistence images, one at each locality, with resolution $20 \times 20$. Convolutions act on the array by comparing the entries of each topological summary from a particular locality with those of its neighbors. This enables the model to learn spatial similarities and multiscale patterns in the topology as the receptive field increases in subsequent convolution layers.

All persistence-based computations are performed using the computational topology library GUDHI \citep{gudhi:urm}. Our pipeline is visually summarized in Figure \ref{fig:pc_pipeline}. In its current form, most of the PHC data generation is a separate preprocessing step applied on the image data outside of the training cycle except for the optimizing the weights of $\mathbf{K}$. Further optimization can be performed on the parameter space which we plan to expand upon in future work, but is outside the scope of this study.

\subsection{Intrinsic Dimensionality of Data}
The manifold hypothesis posits that many high-dimensional datasets are supported on low-dimensional latent manifolds \citep{cayton2005algorithms}. We hypothesize that PHCs reduce the intrinsic dimension of a dataset while retaining important geometric information.

That is, the persistent images are described by significantly fewer latent variables than the original image patches. There are multiple reasons why this might be beneficial:
\begin{itemize}
    \item[1.] Increased memory and resource efficiency, and reduced number of parameters required during training.
    \item[2.] Lower intrinsic dimension is often correlated with better generalization from training to testing sets \citep{pope2021intrinsic}.
    \item[3.] The sample complexity to learn the data manifold depends exponentially on the intrinsic dimension. \citep{narayanan2010sample}.
\end{itemize}   

A commonly used method to estimate the intrinsic dimension of each data representation is the \textit{Maximum Likelihood Estimator} (MLE) proposed by  \citep{levina2004maximum}. The intrinsic dimension is estimated statistically by relating the density of points to the surface area of a spheres of varying dimensions. A complete explanation along with the mathematical construction of these estimators can be found in the references \citep{levina2004maximum, davidcomments}.

\subsection{Model, Training, and Implementation Details} \label{section:implement}

We describe the CNN architecture used for our computational study. Further, our study indicates that a small CNN is sufficient to achieve high performance, consisting only of two convolution and pooling layers (which update the weights of $\mathbf{K}$) followed by three dense layers for the multilayer perceptron. These models are small enough that most modern laptops have sufficient hardware to train them. 

All models are created and trained using TensorFlow \citep{abadi2016tensorflow}. The exact specifications of the architecture is as follows (see \citep{goodfellow2016deep} for an explanation of the terms used below). The kernel size is fixed to $(3,3)$ for both convolutions layers with a stride of two. ReLu is the default activation function used for all layers of the CNN except the last which uses the SoftMax function. The weights of the model are initialized using He Normal Initialization \citep{he2015delving}. We employ Adam \citep{kingma2015adam} for optimization with a fixed learning rate of $\alpha=0.001$. The loss functions used is categorical cross entropy given by the formula \[CE = - \sum_{i=1}^{N}y_i\log(\hat{y}_i)\] where $y_i$ is the true label and $\hat{y}_i$ is the predicted label. Regularizers such as $L_1$, $L_2$, and dropouts are used as well to improve model performance.  These values vary with \[L_1,L_2 \in \{0.0001, 0.001, 0.01, 0.1\}\] and \[\textbf{dropout} \in \{0.1, 0.2, 0.3\} \] Similarly the convolution filter sizes vary from $\{8, 16, 32\}$ and layer size from $\{64, 128, 256, 512, 1028\}$. All of these hyperparameters were chosen using a Bayesian hyperparameter sweep implemented using Weights and Biases \citep{wandb} to maximize for accuracy.  

In addition to comparing model performance between the three distinct representations, we also test the combination of image data with persistence  (images + global PH or images + local PH).  To train a model with the combined representations of image and PH data the CNN model architecture is slightly modified. Instead, it includes two separate feature extraction blocks with the same number of convolution and pooling layers as before using the same sweep configuration listed above. The image and PH data are fed through each separate block and concatenated together before being fed through the dense layers. We hypothesize that the inductive biases (translational invariance, locality, and hierarchical learning \citep{app13095521}) that are built into CNNs enable the model to compare topological signatures in the persistence images across localities to find spatial relationships of local topologies. All five distinct training sets were trained over a 1000 model initializations.

\subsection{Evaluation Metrics}
We evaluate the performance of 1000 models trained on each data type. The hyperparameters of each model are determined using a Bayesian search over the configuration space to optimize the testing accuracy. The image and PHC data is split into training, validation, and testing subsets in a 70/10/20 ratio. Every model is trained over 50 epochs and early stopping is imposed with a patience of $p=5$. Each trained model is also assessed on its precision, sensitivity, and specificity. Testing accuracy is measured by \[\textbf{Accuracy} = \frac{\text{Correct Classifications}}{\text{All Classifications}}. \] 
\,

The latter three metrics are measured in terms of the following number of true positives (TP), true negatives (TN), false negatives (FN), and false positives (FP): 
\[\begin{aligned}\textbf{Precision} &= \frac{TP}{TP + FP} \\
\textbf{Sensitivity} &= \frac{TP}{TP + FN} \\ 
\textbf{Specificity} &= \frac{TN}{TN + FP} .
\end{aligned} \] In binary classification tasks (e.g., distinguish between tumor and non-tumor growth) these metrics have specific interpretations. 

Precision measures the ratio of slides correctly classified with tumor growth against all slides labeled with tumor growth. This metric is used when the consequences of false positives are high (e.g., diagnosis of a terminal disease). 

Sensitivity measures the model's ability to correctly diagnose a patient with tumor growth as positive. Conversely, specificity measures a model's ability classify a patient without the disease as a negative. High sensitivity scores minimize FNs whereas high specificity scores minimize FPs. 
It is important to note that sensitivity and specificity scores are negativity correlated as there is often an overlap between the distribution of diseased and non-diseased populations. For example, a sensitivity score of 100\% can be achieved if a model predicts that every slide has tumor growth, however, this would severally impact the specificity score with numerous false positives. For this reason, sensitivity is computed by predetermining a minimum specificity score and similarly for specificity. We measure sensitivity at a minimum specificity score of $0.9$ and similarly for specificity. 

Precision, sensitivity, and specificity scores are generalized for multi-class classification through aggregation across classes. This is achieved by fixing a specific class as the ``positive" class and all other classes as ``negative" and computing their precision, sensitivity, specificity metrics. This is repeated for each class and the results are averaged to return an aggregated score of each metric.

\section{Results}
For each of the five data representations --- persistent homology convolutions, global persistent homology, truncated images, and combinations thereof --- we chose the model hyperparemeters that resulted in the highest accuracy. A variety of filtration methods were tested to measure the performance of PHC. The full set of results are available in the Appendix \ref{sec:appendix}. Here, we restrict our focus to the two best performing filtrations in dimension one: the alpha complex and extended persistence on the adjacency complex.  Performance metrics for these accuracy-maximizing models are summarized in Table \ref{sec:filtmetrics}. The models trained with local persistent homology the form of PHCs using the alpha complex and extended persistence of the adjacency complex exhibit better performance on all metrics than the models trained with either grayscale images or global persistence. This supports our hypotheses that the local arrangement of topological features is an important characteristic for the differentiation of the three tissue classes, and that at least some of this information is not detected by models trained with a standard CNN architecture.  

Note that in some instances, specific evaluation scores for the alpha complex and the extended adjacency complex are identical. There are two factors that likely contribute to this outcome. First, the repetition in scores is an upper limit of model performance over the hyperparameter sweep; the repetition in scores were present among several of the top performing models across the sweeps. Second, the image dataset from \citep{arunachalam2019viable} are subdivided images of whole slides. In some instances, a subdivision of whole slides yields images with an absence of tissue matter, potentially obfuscating training. This may be reflected in the evaluation score with models repeatedly misclassifying samples.
 
Models trained with adjacency XPHC data tend to yield the best performance across all metrics with the exception of sensitivity and specificity where it either ties with the alpha complex or slightly under-performs. That said, those two filtrations yield very similar performance despite encoding qualitatively different information about the geometry of histopathology slides. Namely, the alpha complex encodes the geometry of the cellular and subcelluar structure whereas the adjacency complex measures the grayscale value of cells in a locality. Moreover, the improved performance of models trained with PHC indicates that additional information encoded --- namely the locality and cell geometry --- are important for histopathological classification in this dataset. 

Table \ref{sec:phc_speed} compares the runtimes required to compute and process PHC and global persistent homology for this dataset across the filtrations. This includes image preprocessing (thresholding/truncating), computing the persistent homology data (either local or global), and vectorizing the resulting data. As expected, persistent homology convolutions can be computed much more quickly than global persistent homology.  

We also consider how model performance varies across the hyperparameter sweep. Table \ref{sec:avg_results} display the average metric performance and standard error of each data type aggregated across all trained models. Models trained with PHC data using any complex have higher average scores and less variance for each performance metric. Interestingly, the adjacency XPHC data has higher average scores in accuracy and precision whereas the alpha complex PHC data has higher average scores in sensitivity and specificity. This indicates that models trained with PHCs on specific filtrations may exhibit less sensitivity to the choice of parameters and may therefore be easier to work with in practice. Moreover, it suggests that PHC can extract meaningful geometric information from the micrographs.

Further evidence for this hypothesis is found by examining the intrinsic dimension estimates of the various data representations presented in Table \ref{sec:data_id}. The intrinsic dimension of the PHC of the data is substantially lower than that of the full images. This indicates that PHC produces a simplified geometric summary that retains information important to the underlying biology. On the other hand, global PH loses even more information and leads to worse performance.  This suggests the importance of making the correct choice of geometric summary, and in particular that the locality of geometric features is significant for histopathological classification in this data set.

\begin{table}[h]

\centering
\renewcommand{\arraystretch}{1.5} 
\scalebox{.85}{
\begin{tabular}{l||c||c||c||c}
\toprule
\textbf{Filtration Type} & \textbf{Accuracy} & \textbf{Precision} & \textbf{Sensitivity} & \textbf{Specificity} \\\midrule
\midrule

Grayscale Image Data & 90.8\% & 91.2\% & 94.3\% & 95.6\%   \\

\hline

 Alpha Complex PHC & 91.7\% & 91.7\% & 94.3\%
 & 95.9\%   \\
\hline

Adjacency Complex XPHC & 91.7\% & 91.7\% & 93.9\& & 95.9\%   \\
\hline

Global Adjacency Complex XPH  & 74.2\% & 76.0\% & 69.4\% & 54.4\%   \\
\hline

Images + Global Adjacency Complex XPH & 91.7\% & 91.7\% & 95.6\% & 96.1\%   \\
\hline
Images + Alpha Complex PHC & 93.5\% & 94.3\% & \textbf{96.5\%} & \textbf{98.0\%}   \\
\hline

Images + Adjacency Complex XPHC & \textbf{93.9\%} & \textbf{94.3\%} & \textbf{96.5\%} & 97.2\%   \\
\hline

\bottomrule
\end{tabular}
} 
\vspace{1em}
\caption{Evaluation metrics for the accuracy-maximizing models of PHC data with varying filtration. The best performing data types for each metric are highlighted in \textbf{bold}. }

\label{sec:filtmetrics}
\end{table}

\begin{table}[h]
\centering
\renewcommand{\arraystretch}{2} 
\begin{tabular}{l||c||c}
\toprule
\textbf{Filtration Type} & \textbf{PHC Data} & \textbf{Global PH Data} \\
\midrule
\midrule
Alpha Complex & \textbf{2.6} sec & 6.8 sec \\
\hline
Extended Adjacency Complex & \textbf{11.4} sec & 3496.0 sec \\
\bottomrule
\end{tabular}
\vspace{1em}
\caption{Average run time to compute persistence on a single image across various complex types on a sample of a $N=100$ images from the Ost. dataset. The dimension of homology computed is fixed to $p=1$. Computations were performed on an M1 MacBook Pro. The fastest compute time is highlighted in \textbf{bold}. }

\label{sec:phc_speed}
\end{table}

\begin{table}[h]
\centering
\renewcommand{\arraystretch}{2} 
\scalebox{1}{
\begin{tabular}{l||c||c}

\toprule
\textbf{Data Representation} & \textbf{Global} & \textbf{Local} \\\midrule
\midrule
 Alpha Complex PH & 0.18 & 4.41 
  \\
  \hline
 Adjacency Complex XPH & 0.18 & 2.24 
  \\
\hline 
Histopathology Images & 24.67 & 12.50  \\
\bottomrule
\end{tabular}
} 
\vspace{1em}
\caption{Intrinsic dimension estimates of various data representations computed using the maximum likelihood estimator proposed by \citep{levina2004maximum, davidcomments}. These results were computed with a sample size of $N=500$. }

\label{sec:data_id}
\end{table}

\begin{table}[t]

\renewcommand{\arraystretch}{1.5} 
\scalebox{.8}{
\begin{tabular}{l||c||c||c||c}
\toprule
\textbf{Data Representation} & \textbf{Accuracy} & \textbf{Precision} & \textbf{Sensitivity} & \textbf{Specificity} \\\midrule
\midrule
Global Adjacency Complex XPH & $63.0 \pm 7.1\%$ & $67.6 \pm 7.6\%$ & $46.9 \pm 10.19\%$ & $46.3 \pm 8.3\%$   \\
\hline
Adjacency Complex XPHC & $84.8 \pm 4.1\%$ & $85.3 \pm 3.9\%$ & $87.2 \pm 6.4\%$ & $86.1 \pm 8.3\%$  \\
\hline

 Alpha Complex PHC & $86.0 \pm 2.2\%$ & $86.2 \pm 2.2$ & $\boldsymbol{89.1 \pm 3.0}$ & $\boldsymbol{88.5 \pm 5.2}$   \\
\hline

Image Data & $80.5 \pm 9.4 \%$ & $81.0 \pm 9.4\%$ & $79.9 \pm 16.7\%$ & $70.6 \pm 30.2\%$   \\
\hline

Images + Alpha Complex PHC & $83.2 \pm 
8.5\%$ & $83.8 \pm 8.1\%$ & $82.9 \pm 17.7\%$ & $76.4 \pm 29.3\%$   \\
\hline

Images + Global Adjacency Complex XPH & $82.5 \pm 8.1\%$ & $82.9 \pm 8.1\%$ & $82.9 \pm 15.1\%$ & $76.6 \pm 27.0\%$   \\
\hline

Images + Adjacency Complex XPHC & $\boldsymbol{86.5 \pm 5.6\%}$ & $\boldsymbol{87.0 \pm 5.2\%}$ & $89.1 \pm 9.6\% $ & $87.5 \pm 16.7\%$   \\

\bottomrule
\end{tabular}
} 
\vspace{1em}
\caption{ Average evaluation metric across all 1000 model initializations. The best performing data types for each metric are highlighted in \textbf{bold}. Observe that the Lowerstar XPHC + Images have the highest averages in accuracy and precision and the TPHC has the highest average sensitivity and specificity scores.}

\label{sec:avg_results}
\end{table}

\section{Conclusion}
\label{sec:conclusion}

We present a novel convolution-like operator, called Persistent Homology Convolutions (PHCs), which augments an image with information representing its local geometry. A comparative study of models trained on an Osteosarcoma dataset demonstrates the effectiveness of PHCs compared to other representations of the data. CNNs trained with PHCs exhibit higher accuracy and less dependence on hyperparameters than conventionally trained neural networks. This suggests that the PHC operation reduces the complexity of the image data to produce a meaningful summary of the the geometry of histopathology slides. For future research, we plan to explore similar operators based on different geometric summaries. Additionally, we hope to further integrate the current PHC generation pipeline with backpropagation to optimally search the parameter space.

\section*{Acknowledgments}
The authors are grateful for the support of the National Science Foundation under Award No. 2232967. Additionally, we would like to thank Ajay Krishna Vajjala, Deval Parikh, and Anthony Pizzimenti for their suggestions regarding the implementation of PHC. The authors would also like to thank Tyrus Berry, Heather Harrington, Kelly Maggs, and Jeremy Mason for their thoughtful comments regarding the content of this paper. Computational resources were provided by the Office of Research Computing at George Mason University. 

\appendix\label{sec:appendix}

\section{Full PHC Experiments}

To evaluate the performance of PHC, we test five filtrations: the ordinary and extended persistence of the alpha complex and the lower star filtration, and the extended persistence of the adjacency complex\footnote{Ordinary persistence was not computed of the adjacency complex as it contains information equivalent to that of the extended persistence. The lower star filtration was computed on a simplicial complex obtained by subdividing the cubical complex formed by the pixels.  Our experimental setup and computational framework is publicly available in our repository: https://github.com/Shrunalp/PHC.git}. We compute the persistent homology of each filtration in dimensions zero and one. For each filtration, we investigate  six training configurations determined by whether the filtration is inputted together with image data or alone, and whether $0$-dimensional features, $1$-dimensional features, or both are included. To accommodate these diverse inputs, the CNN architecture utilizes separate convolution and pooling branches for each data format.

Tables \ref{tab:alpha_complex}-\ref{tab:ext_adj} summarize the accuracy-maximizing models across all filtration. Hybrid models, trained using image and PHC data, consistently outperform the benchmarks of models trained exclusively on image data (shown in Table \ref{sec:filtmetrics}). Furthermore, while extended persistence yields better results than ordinary persistence for the lower star filtration, it underperforms for the alpha complex.

Furthermore, we observe that the best performance was given by the one-dimensional persistent homology features, except for the extended alpha complex. This suggests that the geometry of the cells themselves (e.g. the irregularity of cell shape and size, the presence of multinucleation, etc) is more relevant to tumor classification than the geometry of the empty space between cells.

Observe that models trained on the persistent homology of the adjacency complex perform better than those trained on the persistent homology of the lower star filtration. This is counterintuitive, as the information contained by the former is also encoded in the latter by construction. We hypothesize that the geometric features represented by the adjacency complex are more relevant to the classification than the additional features encoded by the lower star filtration, and that including more information than necessary --- even if it meaningfully corresponds to geometric features --- results in worse model performance.  This could also explain the underperformance of extended persistence of the alpha complex as compared to the ordinary persistence of the same filtration.

Lastly, Table \ref{sec:full_phc_speed} compares the run times of PHC relative to global PH computations across filtrations. As expected, PHC computations are significantly faster.


\begin{table*}[h]
\centering
\renewcommand{\arraystretch}{1.2}
\scalebox{0.85}{
\begin{tabular}{c||c||c|c|c|c}
\toprule
\multirow{2}{*}{\textbf{Alpha Complex}} & \multirow{2}{*}{\textbf{Dimension}} & \multicolumn{4}{c}{\textbf{Model Performance}} \\
\cmidrule{3-6}
& & \textbf{Test Accuracy} & \textbf{Precision} & \textbf{Sensitivity} & \textbf{Specificity} \\
\midrule
\multirow{3}{*}{\text{With Images}}
& 0 & 90.0\% & 90.4\% & 93.9\% & 94.3\% \\
& 1 & \textbf{93.5\%} & \textbf{94.3\%} & \textbf{96.5\%} & \textbf{98.0\%} \\
& 0 + 1 & 91.7\% & 92.8\% & 95.2\% & 96.5\% \\
\midrule
\multirow{3}{*}{\text{Without Images}}
& 0 & 83.8\% & 85.5\% & 86.9\% & 83.2\% \\
& 1 & 91.7\% & 91.7\% & 94.4\% & 95.9\% \\
& 0 + 1 & 87.2\% & 87.6\% & 89.5\% & 87.8\% \\
\bottomrule
\end{tabular}
}
\vspace{1em}
\caption{Evaluation metrics for the accuracy-maximizing models trained using \textbf{Alpha Complex} PHC data. Models are trained across different dimensions in homology, with and without image data.}
\label{tab:alpha_complex}
\end{table*}

\begin{table*}[h]
\centering
\renewcommand{\arraystretch}{1.2}
\scalebox{0.85}{
\begin{tabular}{c||c||c|c|c|c}
\toprule
{\textbf{Extended Persistence}}  & \multirow{2}{*}{\textbf{Dimension}} & \multicolumn{4}{c}{\textbf{Model Performance}} \\
\cmidrule{3-6}
{\textbf{Alpha Complex}} & & \textbf{Test Accuracy} & \textbf{Precision} & \textbf{Sensitivity} & \textbf{Specificity} \\
\midrule
\multirow{3}{*}{\text{With Images}}
& 0 & 88.7\% & 90.0\% & 90.8\% & 91.3\% \\
& 1 & 90.0\% & 89.9\% & 93.0\% & 93.2\% \\
& 0 + 1 & \textbf{91.7\%} & \textbf{91.7\%} & \textbf{95.2\%} & \textbf{96.1\%} \\
\midrule
\multirow{3}{*}{\text{Without Images}}
& 0 & 81.2\% & 83.3\% & 83.0\% & 80.6\% \\
& 1 & 83.8\% & 87.3\% & 87.7\% & 88.9\% \\
& 0 + 1 & 86.0\% & 89.8\% & 87.3\% & 84.9\% \\
\bottomrule
\end{tabular}
}
\vspace{1em}
\caption{Evaluation metrics for the accuracy-maximizing models trained using \textbf{Extended Persistence on the Alpha Complex} PHC data. Models are trained across different dimensions in homology, with and without image data.}
\label{tab:ext_alpha_complex}
\end{table*}

\begin{table*}[h]
\centering
\renewcommand{\arraystretch}{1.2}
\scalebox{0.85}{
\begin{tabular}{c||c||c|c|c|c}
\toprule
{\textbf{Lower Star}}  & \multirow{2}{*}{\textbf{Dimension}} & \multicolumn{4}{c}{\textbf{Model Performance}} \\
\cmidrule{3-6}
{\textbf{Filtration}} & & \textbf{Test Accuracy} & \textbf{Precision} & \textbf{Sensitivity} & \textbf{Specificity} \\
\midrule
\multirow{3}{*}{\text{With Images}}
& 0 & 92.1\% & 92.1\% & 94.3\% & 96.5\% \\
& 1 & \textbf{92.1\%} & \textbf{93.0\%} & \textbf{94.3\%} & \textbf{96.5\%} \\
& 0 + 1 & 90.8\% & 90.8\% & 95.2\% & 95.4\% \\
\midrule
\multirow{3}{*}{\text{Without Images}}
& 0 & 91.3\% & 92.0\% & 95.2\% & 96.1\% \\
& 1 & 88.7\% & 88.6\% & 90.4\% & 90.8\% \\
& 0 + 1 & 90.4\% & 90.3\% & 91.7\% & 95.2\% \\
\bottomrule
\end{tabular}
}
\vspace{1em}
\caption{Evaluation metrics for the accuracy-maximizing models trained using \textbf{Lower Star Filtration} PHC data. Models are trained across different dimensions in homology, with and without image data. }
\label{tab:lower_star}
\end{table*}

\begin{table*}[h]
\centering
\renewcommand{\arraystretch}{1.2}
\scalebox{0.85}{
\begin{tabular}{c||c||c|c|c|c}
\toprule
{\textbf{Extended Persistence}}  & \multirow{2}{*}{\textbf{Dimension}} & \multicolumn{4}{c}{\textbf{Model Performance}} \\
\cmidrule{3-6}
{\textbf{Lower Star}} & & \textbf{Test Accuracy} & \textbf{Precision} & \textbf{Sensitivity} & \textbf{Specificity} \\
\midrule
\multirow{3}{*}{\text{With Images}}
& 0 & 92.1\% & 91.7\% & 94.3\% & \textbf{96.5\%} \\
& 1 & \textbf{92.6\%} & \textbf{92.5\%} & 94.3\% & 96.3\% \\
& 0 + 1 & 90.8\% & 91.2\% & 93.8\% & 96.1\% \\
\midrule
\multirow{3}{*}{\text{Without Images}}
& 0 & 91.2\% & 92.0\% & \textbf{95.2\%} & 96.1\% \\
& 1 & 90.0\% & 90.3\% & \textbf{95.2\%} & 95.0\% \\
& 0 + 1 & 90.3\% & 90.3\% & 91.7\% & 95.1\% \\
\bottomrule
\end{tabular}
}
\vspace{1em}
\caption{Evaluation metrics for the accuracy-maximizing models trained using \textbf{Extended Lower Star Filtration} PHC data. Models are trained across different dimensions in homology, with and without image data. }
\label{tab:ext_lower_star}
\end{table*}

\begin{table*}[h]
\centering
\renewcommand{\arraystretch}{1.2}
\scalebox{0.85}{
\begin{tabular}{c||c||c|c|c|c}
\toprule
{\textbf{Extended Persistence}}  & \multirow{2}{*}{\textbf{Dimension}} & \multicolumn{4}{c}{\textbf{Model Performance}} \\
\cmidrule{3-6}
{\textbf{Adjacency Complex}} & & \textbf{Test Accuracy} & \textbf{Precision} & \textbf{Sensitivity} & \textbf{Specificity} \\
\midrule
\multirow{3}{*}{\text{With Images}}
& 0 & 88.2\% & 88.9\% & 92.1\% & 92.6\% \\
& 1 & \textbf{93.9\%} & \textbf{94.3\%} & \textbf{96.5\%} & \textbf{97.2\%} \\
& 0 + 1 & 91.7\% & 92.8\% & 95.2\% & 96.5\% \\
\midrule
\multirow{3}{*}{\text{Without Images}}
& 0 & 83.0\% & 82.6\% & 83.4\% & 77.1\% \\
& 1 & 91.7\% & 91.7\% & 93.9\% & 95.9\% \\
& 0 + 1 & 90.0\% & 90.0\% & 91.7\% & 94.1\% \\
\bottomrule
\end{tabular}
}
\vspace{1em}
\caption{Evaluation metrics for the accuracy-maximizing models trained using \textbf{Extended Persistence of the Adjacency Complex} PHC data. Models are trained across different dimensions in homology, with and without image data. }
\label{tab:ext_adj}
\end{table*}

\begin{table}[h]
\centering
\renewcommand{\arraystretch}{2} 
\begin{tabular}{l||c||c}
\toprule
\textbf{Filtration Type} & \textbf{PHC Data} & \textbf{Global PH Data} \\
\midrule
\midrule
Alpha Complex & \textbf{2.6} sec & 6.8 sec \\
\hline
Lower Star Filtration & \textbf{5.6} sec & 73.4 sec \\
\hline
Extended Adjacency Complex & \textbf{11.4} sec & 3496.0 sec \\
\hline
Extended Alpha Complex & \textbf{8.1} sec & 10.6 sec \\
\hline
Extended Lower Star Filtration & \textbf{16.2} sec & 92.9 sec \\
\bottomrule
\end{tabular}
\vspace{1em}
\caption{Average run time to compute persistence on a single image across various complex types on a sample of a $N=100$ images from the Ost. dataset. The dimension of homology computed is fixed to $p=1$. Computations were performed on an M1 MacBook Pro. The fastest compute time is highlighted in \textbf{bold}. }

\label{sec:full_phc_speed}
\end{table}

\bibliographystyle{gtml2025_workshop}
\bibliography{gtml2025_workshop}

\end{document}